# Magnetic order induced polarization anomaly of Raman scattering in 2D magnet $CrI_3$

Yujun Zhang[1,2,†], Xiaohua Wu[1,†], BingBing Lyu[1], Minghui Wu[1], Shixuan Zhao[1], Junyang Chen[1], Mengyuan Jia[1], Chusheng Zhang[1], Le Wang[3], Xinwei Wang[2], Yuanzhen Chen[1,3], Jiawei Mei[1,3], Takashi Taniguchi[4], Kenji Watanabe[4], Hugen Yan[5], Qihang Liu[1,3], Li Huang[1], Yue Zhao[1,3,*], Mingyuan Huang[1*]

[1]*Department of Physics, Southern University of Science and Technology, Shenzhen 518055, China*
[2]*School of Advanced Materials, Shenzhen Graduate School Peking University, Shenzhen 518055, China*
[3]*Shenzhen Institute for Quantum Science and Engineering, Southern University of Science and Technology, Shenzhen 518055, China*
[4]*Advanced Materials Laboratory, National Institute for Materials Science, Tsukuba, Ibaraki 305-0044, Japan*
[5]*State Key Laboratory of Applied Surface Physics and Department of Physics, Fudan University, Shanghai 200433, China*

[†]These authors contributed equally.
[*]Correspondence and requests for materials should be addressed to Y. Z. (email: zhaoy@sustech.edu.cn) or M.H. (email: huangmy@sustech.edu.cn)

**Abstract**
The recent discovery of 2D magnets has revealed various intriguing phenomena due to the coupling between spin and other degree of freedoms (such as helical photoluminescence, nonreciprocal SHG). Previous research on the spin-phonon coupling effect mainly focuses on the renormalization of phonon frequency. Here we demonstrate that the Raman polarization selection rules of optical phonons can be greatly modified by the magnetic ordering in 2D magnet $CrI_3$. For monolayer samples, the dominant $A_{1g}$ peak shows abnormally high intensity in the cross polarization channel at low temperature, which is forbidden by the selection rule based on the lattice symmetry. While for bilayer, this peak is absent in the cross polarization channel for the layered antiferromagnetic (AFM) state and reappears when it is tuned to the ferromagnetic (FM) state by an external magnetic field. Our findings shed light on exploring the emergent magneto-optical effects in 2D magnets.

**Introduction**
The discovery of 2D van der Waals magnets has attracted a lot of attention in both fundamental physics and potential applications (*1–9*). Among these 2D magnetic materials, $CrI_3$ is the most intensively studied, where the weak interlayer AFM coupling can be manipulated by magnetic field, electric field, doping and pressure

(*10–15*), leading to many interesting potential applications, such as magnetoresistive memories, spin filters (*16–20*) and spin transistors (*21*). The recent experiment shows that the AFM bilayer $CrI_3$ exhibits giant nonreciprocal second harmonic generation due to the inversion symmetry breaking induced by the layered AFM order (*22*), but how this magnetic ordering affects the inelastic light scattering is still unclear. Moreover, the Raman scattering in monolayer and other thin $CrI_3$ layers with different magnetic ordering is elusive as well.

Raman spectroscopy is a versatile tool to probe magnetic materials and can provide information on energy, symmetry and statistics of magnetic excitations, such as fluctuations of magnetic energy density (quasi-elastic scattering) (*23*, *24*), magnons (*25*, *26*) and spinons (*27*). It has been widely used in the study of magnetic phase transitions (*28*). Besides the magnetic excitations, Raman spectroscopy can be used to study the phonon behavior induced by magnetic order, including spin-phonon coupling (*29*), concomitant structural phase transition (*30*) and Brillouin zone folding (*9*, *31*). In the study of recent 2D magnets, Raman spectroscopy provided the first evidence of long range AFM order in monolayer $FePS_3$ (*9*) and was subsequently used to probe the magnetic phase transition in $Cr_2Ge_2Te_6$ (*30*) and $CrI_3$ (*32*).

In this work, we study the magneto-optical effect of $CrI_3$ thin layers by probing the phonon properties through polarized Raman spectroscopy. For monolayer, the dominant Raman mode at about 128 $cm^{-1}$ displays abnormally high intensity in the cross polarization channel at low temperature. Moreover, an abrupt intensity change in both parallel (XX) and cross (XY) polarization channels is observed with the spin flip driven by the external magnetic field. However, this peak (at about 129 $cm^{-1}$ for bilayer) is absent in the XY channel for the AFM bilayer and reappears when the bilayer is tuned to FM state by a magnetic field. A new peak at around 127 $cm^{-1}$, presumably due to the Davydov splitting, can be observed in the AFM bilayer, though it's silent in the FM state due to the inversion symmetry changes. This anomalous polarization behavior cannot be explained by the lattice symmetry and the magnetic ordering must have been taken into account. By group theory analysis, we highlight that the polarization selection rule of optical phonons can be greatly modified by the magnetic order. Furthermore, we also demonstrate that Raman spectroscopy can probe the magnetic order transitions in 3 and 4-layer $CrI_3$ and this polarization anomaly gets weaker as the number of layer increases.

**Results**

We first investigated the Raman spectrum of $CrI_3$ monolayers as a function of temperature. Representative Raman spectra of a monolayer sample at 1.7 K are displayed in Fig. 1A, with four Raman modes labeled as $A_1$, $A_2$, $E_1$ and $E_2$, similar to previous studies (*32–36*). The dominant Raman peak $A_2$ only shows up in the XX channel above the critical temperature ($T_C$ about 45 K) (*1*), as shown in supplementary fig. S1 (see details in the supplementary note 1). After the sample is cooled down below $T_C$, this $A_2$ peak starts to show up in the XY channel and its intensity is even higher than that of the XX channel at 1.7 K. Such polarization change indicates strong coupling between this phonon and the magnetic order. Note that this work mainly focuses on the behavior of this $A_2$ mode.

We next explored the effect of applied out-of-plane magnetic field. Figure 1(B to E) show the evolution of the Raman spectra of the monolayer sample from both XX and XY channels with increasing (Fig. 1, B and D) and decreasing (Fig. 1, C and E) magnetic field. We can clearly see a sudden increase (decrease) of the intensities in both XX and XY channels at around 0.075 T (-0.075 T) with increasing (decreasing) magnetic field. We plot the total intensity (from both XX and XY channels) versus the magnetic field together with the reflective magnetic circular dichroism (RMCD) signal in Fig. 1F, where the observed hysteresis loop is a direct evidence showing the spin flip driven by the external magnetic field. The spin direction dependent Raman scattering may result from the resonant effect by using 633 nm laser (*15*). The coercive field extracted from the Raman hysteresis loop (about 0.075 T) is less than that from the RMCD measurement (about 0.2 T), which is probably due to the higher laser power used in our Raman measurement (see details in materials and methods).

The crystal structure of monolayer $CrI_3$ belongs to $D_{3d}$ point group and the $A_2$ peak is assigned as an $A_{1g}$ mode, associated with the symmetrical breathing mode between the upper and lower Iodine layers as shown in the inset of Fig. 1A (*34*). From the selection rule based on lattice symmetry, the scattering signal should only appear in the XX channel. One possible reason for the appearance of the Raman signal in the XY channel is a new Raman mode with the same frequency as the $A_2$ mode, such as magnon (*32*). We performed Raman measurement on the monolayer sample at high magnetic field up to 8 T. However, no observable shift or splitting was detected for this mode, as shown in supplementary fig. S2, suggesting a different origin. Given the fact that the intensities from the XX and XY channels show similar behaviors under a magnetic field, we conclude that they are associated with the same phonon mode. This assignment can be further verified by the group theory analysis.

Based on the group theory, below we demonstrate that the polarization selection rules of optical phonons can be modified when the crystal undergoes a transition from a paramagnetic to a magnetic state as the time reversal symmetry is broken. The monolayer $CrI_3$ has the crystallographic symmetry $D_{3d} = \{e, 2C_3, 3C_2', i, 2S_6, 3\sigma_d\}$, and the Raman tensor of the $A_2$ mode is derived as equation 1a. If there is out-of-plane FM spin alignment, the corresponding magnetic point group is $D_{3d}(S_6) = \{e, 2C_3, 3RC_2', i, 2S_6, 3R\sigma_d\}$, where $R$ is the time reversal operator. According to the group theory, the Raman tensor of the optical phonon in the magnetic ordered state will have the same form as that of the optical magnon because they are constrained by the same symmetry. Therefore, the Raman tensor of this phonon mode will turn into equation 1b in the FM state (*37, 38*).

$$A_{1g} = \begin{pmatrix} a & 0 & 0 \\ 0 & a & 0 \\ 0 & 0 & c \end{pmatrix} \quad (1a) \qquad A_g = \begin{pmatrix} a & ib & 0 \\ -ib & a & 0 \\ 0 & 0 & c \end{pmatrix} \quad (1b) \qquad 1$$

where *a-c* are constants and $i$ is $\sqrt{-1}$. Obviously, the imaginary antisymmetric terms in the Raman tensor caused by the magnetic order can account for the Raman scattering signal observed in the XY channel (see details in the supplementary note 3).

For the CrI$_3$ samples with two and more layers, we observed a new peak at the lower frequency side of the A$_2$ peak at the temperature below $T_C$, which is consistent with a previous report (*32*). Figure 2 (A and B) show the evolution of the Raman spectra of a bilayer sample in the XX and XY channels respectively with increasing magnetic field from -1.5 to 1.5 T. In the XY channel, we observe that the original A$_2$ peak (129 cm$^{-1}$) disappears and the split peak at about 127 cm$^{-1}$ emerges at B = -0.65 T. As the magnetic field increases, the situation is reversed at B = 0.65 T. As shown in Fig. 2D, RMCD plots indicate that magnetic phase transitions occur at B = -0.65 T (from spin down FM state to AFM state) and 0.65 T (from AFM state to spin up FM state), suggesting a correlation between the Raman polarization change and the magnetic phase transition. In the XX channel, only the original A$_2$ peak can be observed and a clear intensity change can be resolved at the critical magnetic fields. We also performed the same measurement with decreasing magnetic field, as shown in supplementary fig. S4. No hysteresis is observed, which agrees with an AFM ground state of bilayer CrI$_3$.

Figure 2C shows the Raman spectra in the XY channel under the magnetic field -1.0, 0 and 1.0 T. The appearance of Raman signal for the A$_2$ mode in both XX and XY channel when the bilayer is tuned to the FM state is similar to the behavior of the monolayer. It is also worth noting that the Raman intensity from the XY channel at 1.0 T is higher than that of the spectrum at -1.0 T, similar to the Raman intensity behavior observed on monolayer as well. By plotting the ratio of the total intensity between the split and the original A$_2$ peaks (Fig. 2D), the magnetic phase transition can be revealed clearly by Raman, as compared to RMCD measurement.

Recent theoretical and experimental studies show that the thin layer CrI$_3$ has the structure of monoclinic phase (space group *C2/m*) other than rhombohedral phase (space group $R\bar{3}$) for bulk crystal at low temperature (*13, 14, 39, 40*), which contributes to the interlayer AFM coupling. Because the lateral interlayer translation breaks most of the symmetries in monolayer, the CrI$_3$ bilayer only has the symmetry $C_{2h}$ = {*e*, $C_2$, *i*, $\sigma_h$}. When the bilayer is at AFM (FM) state, the corresponding magnetic point group is $C_{2h}(C_2)$ = {*e*, $C_2$, $Ri$, $R\sigma_h$} ($C_{2h}(C_i)$ = {*e*, $RC_2$, *i*, $R\sigma_h$}), and the Raman tensor of this phonon mode has the form of equation 2a (2b) (*37, 38*).

$$A = \begin{pmatrix} a & 0 & d \\ 0 & f & 0 \\ h & 0 & c \end{pmatrix} \quad (2a) \qquad A_g = \begin{pmatrix} a & ib & d \\ ie & f & ig \\ h & ik & c \end{pmatrix} \quad (2b) \qquad 2$$

where *a-h* and *k* are constants. For Van der Waals materials, the Raman spectra are mainly determined by their monolayer structure because the Van de Waals interaction between layers is weak and the lattice stacking order only has a small impact on the Raman spectrum. As a consequence, we have $f \approx a$, $e \approx -b$ and $d, h, g, k \approx 0$ and Raman tensor 2a (2b) turns into 1a (1b) (see details in the supplementary note 5). Consequently, for the AFM bilayer, the Raman scattering in the XY channel is prohibited, but allowed for the FM bilayer, which is fully consistent with our experimental observation.

In comparison to the monolayer, there exists interlayer coupling in multi-layer CrI$_3$, thus the original $A_{1g}$ mode of the monolayer splits into *N* modes in *N*-layer CrI$_3$.

Such splitting is termed as Davydov splitting (*41, 42*). Given the facts of the spectrum weight transfer behavior and small frequency difference between the split and original $A_2$ modes, we suggest that the split new peak at 127 cm$^{-1}$ is originated from the Davydov splitting. To confirm our assignment, we performed the first principles calculation within density functional theory on bilayer CrI$_3$ to determine the frequencies of its vibrational modes (see details in the supplementary note 6). The calculated frequency difference between those two modes originated from the $A_2$ mode is about 1.8 cm$^{-1}$, which agrees very well with our experimental value of 2.0 cm$^{-1}$. From the symmetry analysis, the $A_2$ mode ($A_{1g}$) splits into an $A_g$ (Raman active) and a $B_u$ (Raman inactive) mode in bilayer CrI$_3$. With a layered AFM spin order, the $B_u$ mode turns into a Raman active $B$ mode due to the inversion symmetry breaking; as of the FM spin order, on the other hand, this mode turns into a Raman inactive $A_u$ mode. Such mode assignments are in good agreement with our experimental findings that the split peak can only be observed in the AFM phase. Assuming a weak interlayer coupling, we can predict that this split peak can only show up in the XY channel (see details in the supplementary note 5).

The field-dependent measurements demonstrate that the Raman spectra of the monolayer and bilayer CrI$_3$ strongly depend on their magnetic ordering, making Raman spectroscopy a sensitive probe for layered magnetism. We then apply Raman technique to investigate CrI$_3$ with increasing number of layers. Figure 3 (A and B) show the Raman spectra from the XY channel for a trilayer (3L) and a four–layer (4L) sample at B = -2.0, -1.0, 0 (for 3L, two different spectra for increasing and decreasing magnetic field ), 1.0 and 2.0 T, along with the RMCD plots in the insets. For both samples, the split peak at around 126 cm$^{-1}$ is suppressed at the fully polarized FM state and the intensity ratio between the split and original peaks changes with interlayer spin configurations. The 3L sample shows different Raman spectra at B = 0 T for the increasing and decreasing magnetic field, originating from its FM ground states. Moreover, no hysteresis is observed for the 4L sample.

According to the previous group theory analysis, when the multilayer CrI$_3$ is tuned to the FM state, the polarization selection rule of the $A_2$ Raman mode is the same as that of the monolayer and the magnetic order induces Raman scattering signal in the XY channel. To investigate the dimensional effect of this behavior, we plot the Raman spectra of fully polarized CrI$_3$ samples with different layers in both XX and XY channels in Fig. 4. The Raman intensity ratio of this $A_2$ mode between the XY and XX channel is extracted as a function of number of layers in the inset. Assuming that the magnetic ordering only contributes the Raman scattering signal in the XY channel, this intensity ratio decreases with the increasing layer number, which may indicate that this effect is related to the thermal fluctuation.

**Discussion**

In summary, we demonstrate that the Raman spectrum depends strongly on the magnetic order in thin layer CrI$_3$. The polarization properties of the Raman peak associated with the layer breathing mode cannot be explained by the lattice symmetry and the magnetic order induced symmetry changes has to be considered. The sensitivity of the Raman spectrum to the magnetic order implies strong spin-lattice coupling in CrI$_3$ thin layer, which results in anomalous polarization properties other than the renormalization of the phonon energy. In addition, this polarization anomaly is getting stronger with thinner CrI$_3$ sample, which may indicate that the spin-lattice coupling is stronger in 2D magnets. Similar effect could be observed in other 2D magnetic materials.

**Materials and Methods**
**Sample fabrications**
Bulk CrI$_3$ crystals were grown by chemical vapor transport method by using the recipe described in reference (*43*). Thin layer CrI$_3$ samples were mechanically exfoliated on Si substrates with 285 nm thick SiO$_2$ layer. The flakes were then encapsulated by hBN flakes using a polymer-assisted transfer technique (*11*). All the above mentioned fabrication processes were carried out inside a glove box with Ar atmosphere to avoid sample degradation by moisture and oxygen. The numbers of layers were initially estimated by optical contrast and later confirmed by both the atomic force microscopy (AFM) measurements and RMCD measurements.

**Reflective magnetic circular dichroism (RMCD)**
RMCD was performed in an autoDRY 2100 cryostat with a base temperature of 1.7 K. A HeNe laser at 633 nm was coupled to the cryostat by using free space optics. The laser beam was first tuned by a chopper with frequency of 800 Hz and then passed through a photoelastic modulator (PEM), which operated at 50.2 KHz with a maximum retardance of $\lambda/4$. The modulated laser beam was focused onto the sample by a low temperature objective with power less than 5 $\mu$W and the reflected light was collected by the same objective and detected by a photodiode. The RMCD signal was given by the ratio of ac component at 50.2 KHz and 800 Hz (both measured by lock-in amplifiers).

**Polarized Raman spectroscopy**
Polarized Raman spectroscopy was performed using a home-built microscope system in the back-scattering geometry. A HeNe laser at 633 nm was used as the excitation laser and the power was kept below 80 $\mu$W to avoid potential sample damage. The laser beam was focused onto the CrI$_3$ samples along the out-of-plane direction by a 40X low temperature objective. The low temperature and magnetic field was provided by an AttoCube closed-cycle cryostat (autoDRY 2100). The scattering light was collected by the same objective and detected by a spectrometer with an 1800 groove/mm grating and a TE-cooled CCD. The excitation light passed through a linear polarizer and a second polarizer was used for the scattered light to select the component either parallel (XX channel) or perpendicular (XY channel) with respect to the incident light polarization. For the measurement under a magnetic field, the polarizer on the scattered light beam was adjusted to compensate the Faraday rotation angle caused by the low temperature objective.

**Supplementary Materials**

Supplementary material for this article is available at……

Note 1: Temperature dependence of Raman spectrum of monolayer $CrI_3$

Note 2: Raman spectrum of monolayer $CrI_3$ under high magnetic field

Note 3: The polarization of the $A_2$ mode in monolayer $CrI_3$

Note 4: The polarized Raman spectra for bilayer $CrI_3$ at decreasing magnetic field

Note 5: Extended group theory analysis for bilayer $CrI_3$

Note 6: Density functional theory calculation of Davydov splitting for bilayer $CrI_3$

Fig. S1. Temperature dependent Raman spectra in both XX and XY channels for monolayer $CrI_3$.

Fig. S2. Raman spectra of monolayer $CrI_3$ under higher magnetic field.

Fig. S3. The polarization analysis of the $A_2$ mode for monolayer $CrI_3$.

Fig. S4. The polarized Raman spectra for bilayer $CrI_3$ at decreasing magnetic field.

Fig. S5. Density functional theory calculation of Davydov splitting for bilayer $CrI_3$.

**Acknowledgements**
**General**: We acknowledge the helpful discussions with Wang Yao. **Funding:** This work was supported by the Science, Technology and Innovation Commission of Shenzhen Municipality (Grant Nos. JCYJ20170412152334605 and JCYJ20160613160524999), National Natural Science Foundation of China (No. 11674150), Guangdong Innovative and Entrepreneurial Research Team Program (No. 2016ZT06D348), the Key R&D Program of Guangdong Province (No. 2019B010931001), Bureau of Industry and Information Technology of Shenzhen (No. 201901161512). The computer time was supported by the Center for Computational Science and Engineering of Southern University of Science and Technology. **Author contributions:** M.H. and Y.Z. conceived and initiated this project; L.W. and J.M. synthesized and characterized the bulk $CrI_3$ single crystals; T.T. and K.W. provided hBN crystals; S.Z., Q.L. and M.H. performed the group theory analysis; M.W. and L.H. conducted the DFT calculation; X.W., B.L., J.C. and C.Z. fabricated and characterized the few-layer samples; Y.Z. and M.J. performed the Raman and RMCD measurements; M.H., Y.Z. H.Y. Y.C. and X.W. analyzed the data; M.H. and Y.Z. wrote the paper and all authors participated in the discussions of the results. **Competing interests:** The authors declare that they have no competing interests. **Data availability:** The data that support the plots of this paper are available from the corresponding authors upon reasonable request.

**Figure captions**

**Fig. 1. Magnetic field dependence of the polarized Raman spectra for monolayer (1L) CrI$_3$.** (**A**) Raman spectra of a monolayer CrI$_3$ sample in the parallel (XX, red) and cross-polarization channels (XY, blue) at low temperature (1.7K). The phonon modes are labeled as A$_1$, E$_1$, E$_2$ and A$_2$. The inset is the atomic structure of monolayer CrI$_3$, where Cr$^{3+}$ is red and I$^-$ is cyan, with the arrows indicating the vibrational mode associated with the A$_2$ peak. (**B** to **E**) Color plots of polarized Raman spectral intensity versus Raman shift and magnetic field as for monolayer CrI$_3$ in the XX (B and C) and XY channels (D and E), where the red arrows indicate the B field sweep direction: increasing (B and D) and decreasing (C and E) magnetic field. (**F**) Reflective magnetic circular dichroism (RMCD) (red) and the total intensity (from both XX and XY channels, blue) over a cycle of the magnetic field. Hysteresis can be observed for both RMCD and Raman intensity.

**Fig. 2. Magnetic field dependence of the polarized Raman spectra for bilayer (2L) CrI$_3$.** (**A** and **B**) Color plots of the polarized Raman spectral intensity versus Raman shift and increasing magnetic field for bilayer CrI$_3$ in the XX and XY channels, respectively. In the XX channel, only the original A$_2$ peak at 129 cm$^{-1}$ can be resolved and its intensity undergoes an abrupt change during the magnetic phase transition between the AFM and FM states; in the XY channel, the A$_2$ peak can be observed at the FM state, but disappears at the AFM state, and the situation is reversed for the split peak at 127 cm$^{-1}$. (**C**) Raman spectra in the XY channel at B = -1.0, 0 and 1.0 T: line-cut profiles indicated by the purple, green and red lines in (B), respectively. (**D**) RMCD (red) and the ratio of the total intensity (blue) between the split and original A$_2$ peaks over a cycle of the magnetic field.

**Fig. 3. Probing the magnetic order in trilayer (3L) and four-layer (4L) CrI$_3$ by Raman spectroscopy.** (**A** and **B**) Raman spectra in the XY channel for 3L CrI$_3$ (A) and 4L CrI$_3$ (B) at different magnetic fields, where the magnetic order can be deduced from the RMCD plot (insets of (A) and (B)). The split peak at around 126 cm$^{-1}$ is suppressed at the fully polarized FM state and the intensity ratio between the split and original A$_2$ peaks changes with interlayer magnetic order.

**Fig. 4. Layer dependence of the polarized Raman spectra in the XX (red) and XY channel (blue) for 1-5 L CrI$_3$ at spin-down FM state.** The inset shows the Raman intensity ratio of the A$_2$ peak between the XY and XX channels as a function of the layer number. The intensity ratio decreases monotonically with the number of layers.

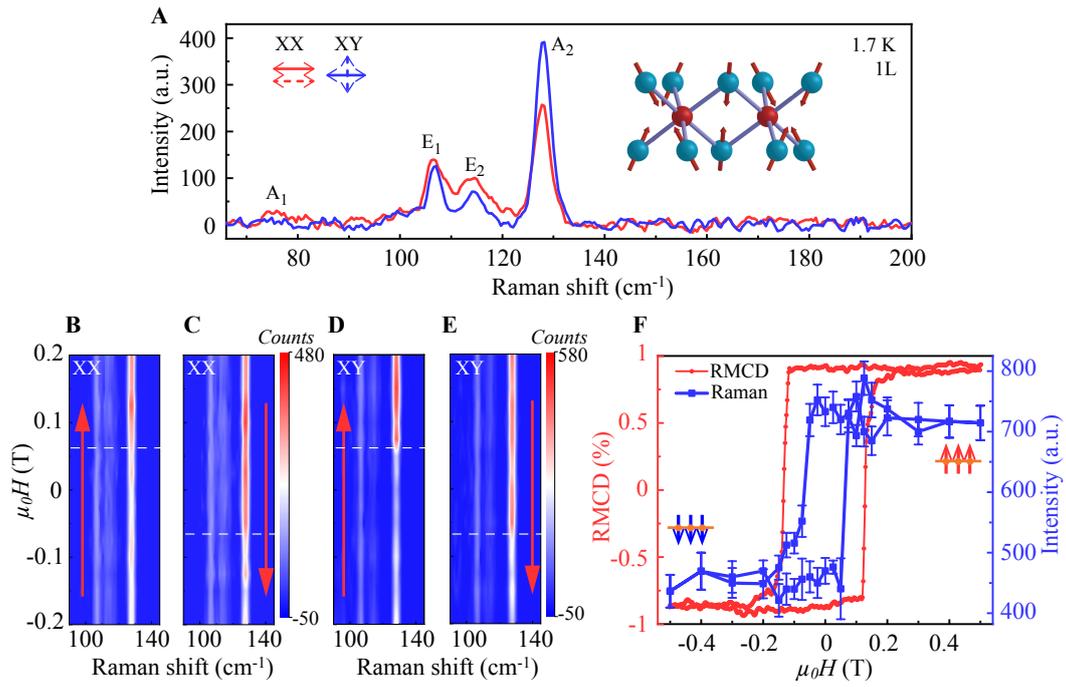

Figure 1

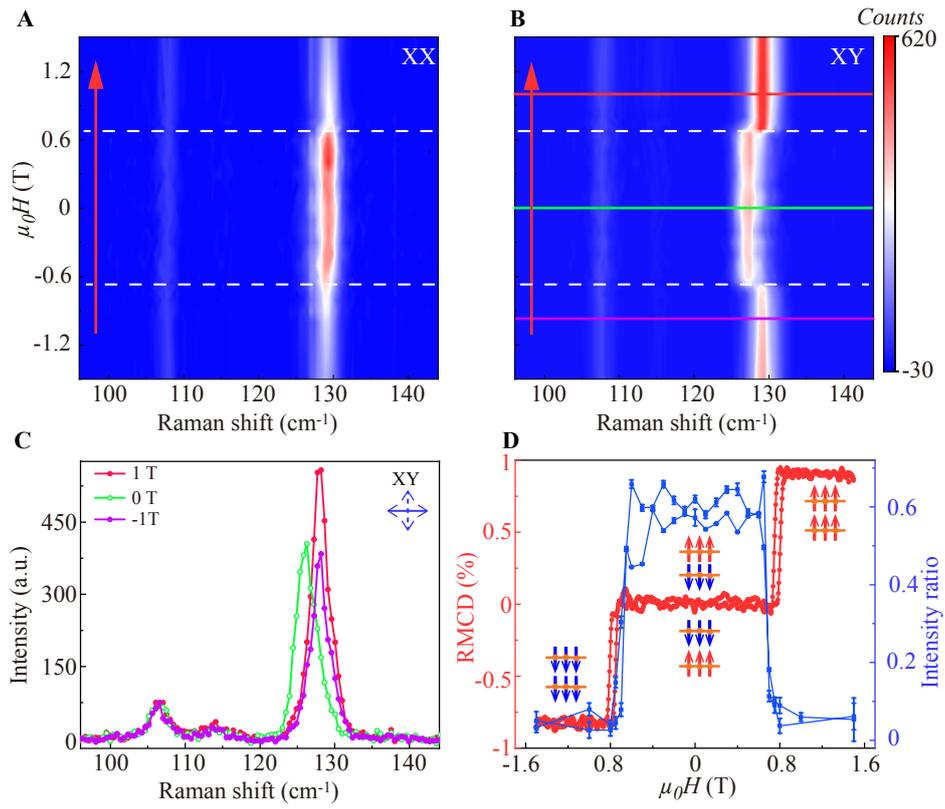

**Figure 2**

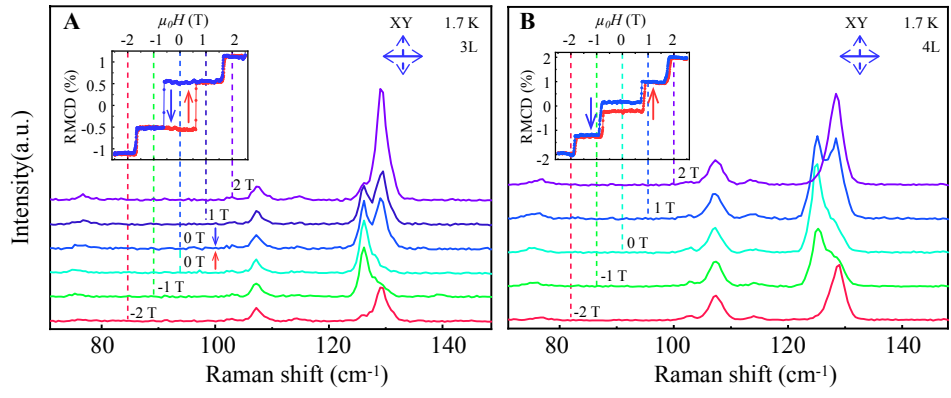

**Figure 3**

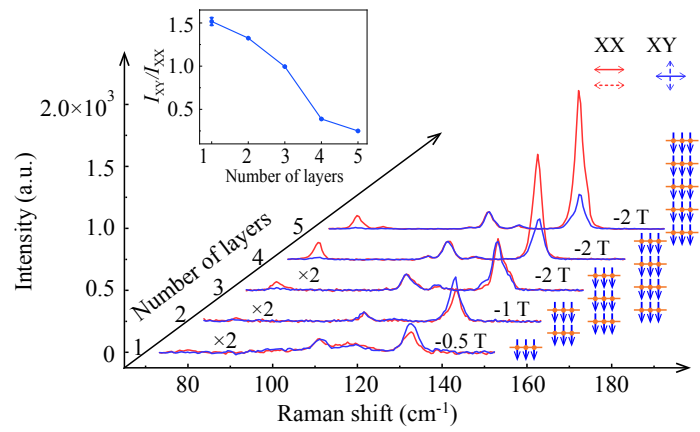

**Figure 4**